\documentclass{aa}  
\usepackage{natbib}

\usepackage{graphicx}
\usepackage{txfonts}
\begin{document}

   \title{Stellar Systems in the direction of the Hickson Compact Group 44 - I. Low Surface Brightness Galaxies}

   \author{A. V. Smith Castelli,
          \inst{1,2}
          F. R. Faifer,\inst{1,2,3}
	  C. G. Escudero\inst{1,2,3}
          }

   \institute{Consejo Nacional de Investigaciones Cient\'ificas y T\'ecnicas, 
              Godoy Cruz 2290, C1425FQB, CABA, Argentina\\  
         \and
             Instituto de Astrof\'isica de La Plata, UNLP, CONICET,  
             Facultad de Ciencias Astron\'omicas y Geof\'isicas, Paseo 
             del Bosque s/n, B1900FWA, La Plata, Argentina\\
             \email{asmith@fcaglp.unlp.edu.ar}\\
	\and
		Facultad de Ciencias Astron\'omicas y Geof\'{\i}sicas,
      Universidad Nacional de La Plata,
      Paseo del Bosque s/n, B1900FWA, La Plata,
      Argentina\\
}

   \date{Received September 15, 1996; accepted March 16, 1997}

 
  \abstract
   {In spite of the numerous studies of low-luminosity galaxies in different environments, 
there is still no consensus about their formation scenario. In particular, a large number of 
galaxies displaying extremely low-surface brightnesses have been detected in the last year, 
and the nature of these objects is under discussion. }
   {In this paper we report the detection of two extended low-surface brightness (LSB)  
 objects ($\mu_{\rm eff_{g'}}\simeq27$ mag) found, in projection, next to NGC\,3193 and in 
the zone of the Hickson Compact Group (HCG) 44, respectively.}
   {We analyzed deep, high-quality, GEMINI-GMOS images with ELLIPSE within IRAF in order to
obtain their brightness profiles and structural parameters. We also search for the presence of globular clusters (GC) in these fields.}
   {We have found that, if these LSB galaxies were at the distances of NGC\,3193 and 
HCG\,44, they would show sizes and luminosities similar to 
those of the ultra-diffuse galaxies (UDGs) found in the Coma cluster and other 
associations. In that case, their sizes would be rather larger than those 
displayed by the Local Group dwarf spheroidal (dSph) galaxies. We have detected
a few unresolved sources in the sky zone occupied by these galaxies showing 
colors and brightnesses typical of blue globular clusters.}
   {From the comparison of the properties of the galaxies 
presented in this work, with those of similar objects reported in the literature, we have
found that LSB galaxies display sizes covering a quite extended continous 
range ($r_{\rm eff}\sim 0.3-4.5$ kpc), in contrast to {\it ``normal''}
early-type galaxies, which shows $r_{\rm eff}\sim 1.0$ kpc with a low dispersion. This fact 
might be pointing to different formation processes for both types of galaxies.}

   \keywords{methods: observational -- techniques: photometric -- galaxies: groups: individual: HCG\,44 --
galaxies: star clusters: general -- galaxies: dwarfs 
               }

\authorrunning{Smith Castelli et al.}
\titlerunning {Stellar Systems in the direction of HCG\,44 - I. LSB galaxies}
   \maketitle
%

\section{Introduction}
\label{introduccion}

Despite the numerous studies of low-mass galaxies performed through 
observations and numerical simulations, their formation scenario is still 
hotly debated. As an example, in the case of low-mass early-type galaxies
located in rich groups and clusters, several authors have claimed that they 
might be gas-rich disk galaxies entering the cluster and
being transformed through the interaction with the 
intracluster medium (e.g. \citealp{2009ApJS..182..216K,2014ApJ...786..105J,2015ApJ...799..172T}). 
Some others, however, based on the strong photometric relations defined
by them, propose a scenario
of in situ formation within the cluster environment 
\citep{2013ApJ...772...68S,2015MNRAS.454.2502S}. In addition, 
there is observational and theoretical evidence that support a third scenario
in which low-mass galaxies might arise 
from the interaction of massive gas-rich galaxies, as bound structures in the 
filaments or bridges that are formed as a consecuence of 
the encounters (e.g. \citealp{1996ApJ...462...50H,2013MNRAS.429.1858D,2014MNRAS.440.1458D}).

The very faint end of the early-type galaxy population is defined 
by dwarf spheroidal (dSph) galaxies, 
which are extended objects displaying
extremely low-surface brightnesses and no evidence of star formation. 
In the last decade, the number of studies of low-surface brightness 
(LSB) galaxies outside 
the Local Group (LG) has increased significantly, thanks to the 
development of detection surveys with amateur telescopes 
(e.g. \citealp{2016A&A...588A..89J}), and the access 
to 8-m class telescopes capable to obtain extremely high-quality
deep images that allow to follow up the objects identified with small
telescopes (e.g. \citealp{2016MNRAS.457L.103R}). The interest in the
identification of new examples of such extremely faint galaxies resides in 
the fact that they can be used as test-beds for constraining models 
predictions. As an example,
$\Lambda$-Cold Dark Matter ($\Lambda$CDM) models produce a larger number 
of dSph satellites around bright galaxies than observed (the so-called Missing
Satellite Problem, e.g., \citealp{1999ApJ...522...82K,1999ApJ...524L..19M}), 
as well as an
isotropic distribution around the brightest galaxies which is not
detected in the LG and M\,81 \citep{2013MNRAS.435.1928P,2013AJ....146..126C}. 
In addition, hierarchical models predict
DM dominated dSph galaxies, while faint galaxies of tidal origin would not 
contain DM at all (\citealp{2013MNRAS.429.1858D} and references therein). 

HCG\,44 was originally classified as a Hickson Compact Group dominated by 
an elliptical galaxy (NGC\,3193 or HCG44b), with three additional 
bright galaxy members: 
an Sa galaxy (NGC\,3189/3190 or HCG44a), an SBc galaxy (NGC\,3185 or 
HCG44c) and an Sd galaxy (NGC\,3187 or HCG44d) 
\citep*{1982ApJ...255..382H,1989ApJS...70..687H}. 
Later, \citet{1991AJ....101.1957W}
identified the dwarf like galaxy [WMv91]\,1015+2203 at the redshift of 
the group, and \citet*{2010ApJ...710..385B} detected four additional members
in their HI study of the group, increasing the HCG\,44 galaxy population
to nine members.
However, \citet{2001ApJ...546..681T} obtained a surface brightness fluctuation 
(SBF) distance of 34 Mpc to NGC\,3193 and, more recently, 
\citet{2013MNRAS.428..370S} have considered [WMv91]\,1015+2203 as a background 
galaxy due to its radial 
velocity ($V_{\rm r}=$1940 km s$^{-1}$), larger than that of the group 
($V_{\rm r}=$1379 km s$^{-1}$). In addition, in their spectroscopic
study of the AGN activity in Hickson Compact Groups, 
\citet{2010AJ....139.1199M} reconsidered NGC\,3193 as a member of HCG\,44. 
Therefore, the real galaxy 
content of HCG\,44 is still under discussion.

In this paper we present a photometric study of two
low-surface brightness galaxies detected in the zone of HCG\,44. 
One is located, in projection, within the 
halo of NGC\,3193, and the other, between the galaxies NGC\,3189/3190
and NGC\,3185. We analyze their structural properties considering
the distance modulii reported in the literature
for the bright galaxies placed in this region of the sky 
(Table\,\ref{distancias}), without attempting 
to clarify the membership status of the latters to the HCG\,44. 
The paper is 
organized as follows. In Section\,\ref{observaciones} we describe the 
photometric data, in Section\,\ref{resultados} we present
the results obtained from the analysis of these data, in
Section\,\ref{Globulars} we analyze the existence of globular
clusters associated with these galaxies, and in 
Section\,\ref{conclusions}, we present  
our conclusions.

\begin{table*}
\caption{Distance information of the bright galaxies present in the region of the HCG\,44.}
\label{distancias}
\centering
\begin{tabular}{lccccccc}
\hline\hline
\multicolumn{1}{c}{Galaxy} & \multicolumn{1}{c}{R.A.} & \multicolumn{1}{c}{DEC} & \multicolumn{1}{c}{(m-M)} & \multicolumn{1}{c}{Distance} & \multicolumn{1}{c}{Method} & \multicolumn{1}{c}{$\rm H_0$} & \multicolumn{1}{c}{Reference} \\
\multicolumn{1}{c}{} & \multicolumn{1}{c}{(J2000)} &\multicolumn{1}{c}{(J2000)} & \multicolumn{1}{c}{(mag)} & \multicolumn{1}{c}{(Mpc)} & \multicolumn{1}{c}{} & \multicolumn{1}{c}{(km s$^{-1}$ Mpc$^{-1}$)}  & \multicolumn{1}{c}{}  \\
\hline
NGC\,3185 	& 10:17:38.5 & 21:41:18 & 32.00  & 25.1 & TF    & 74.4 & (1)  \\
NGC\,3187 	& 10:17:47.8 & 21:52:24	& 32.08  & 26.1 & T est & 75.0 & (2)  \\
          	& 	     & 		& 32.62  & 33.3 & TF    &      & (3)  \\
NGC\,3189/3190 	& 10:18:05.6 & 21:49:56 & 31.73  & 22.2 & SNI   & 74.4 & (1)  \\
NGC\,3193 	& 10:18:24.9 & 21:53:38 & 32.63  & 33.6 & SBF   & 74.4 & (1) \\
\hline
\end{tabular}
\tablefoot{TF: Tully-Fisher; T est: Tully estimation; SNI: Supernova Type I; SBF: Surface Brightness Fluctuations; 
(1): \citet{2013AJ....146...86T}; (2): Nearby Galaxy Cataloge (1988, Tully R. B.); (3): \citet{2007A&A...465...71T}.
}
\end{table*}


\section{Observational Data}
\label{observaciones}

In semester 2010B we have obtained, with the 
Gemini Multi-Object Spectrograph (GMOS) \citep{2004PASP..116..425H} of 
GEMINI-North (Program: GN-2010B-Q-29, PI: A. Smith Castelli),  $g'$, $r'$, $i'$ and $z'$
images of NGC\,3193, and of one field between 
NGC\,3189/3190 and NGC\,3185, in the zone of the HCG\,44. 
The instrument consists 
of 3 CCDs of 2048 $\times$ 4096 pixels, separated by gaps of $\sim$ 2.8 arcsec, 
with an unbinned pixel scale of 0.0727 arcsec pixel$^{-1}$.  The
field of view (FOV) is 5.5' $\times$ 5.5' and the scale for binning 2$\times$2 
is 0.146 arcsec pixel$^{-1}$. In Table\,\ref{info} we show the information related 
to the GEMINI-GMOS images.

\begin{table*}
\caption{Log of the GEMINI-GMOS images.}
\label{info}
\centering
\begin{tabular}{ccccccc}
\hline\hline
\multicolumn{1}{c}{} & \multicolumn{3}{c}{NGC\,3193} & \multicolumn{3}{c}{HCG\,44}\\
\hline
\multicolumn{1}{c}{Filter} & \multicolumn{1}{c}{Date} & \multicolumn{1}{c}{Exposures} & \multicolumn{1}{c}{FWHM} & \multicolumn{1}{c}{Date} & \multicolumn{1}{c}{Exposures} & \multicolumn{1}{c}{FWHM}\\
\multicolumn{1}{c}{} & \multicolumn{1}{c}{} & \multicolumn{1}{c}{(sec)} & \multicolumn{1}{c}{(arcsec)}  & \multicolumn{1}{c}{}  & \multicolumn{1}{c}{(sec)} & \multicolumn{1}{c}{(arcsec)}\\
\hline
g' & 17 Nov. 2010 & 5$\times$340 & 0.8 & 05 Jan. 2011 & 4$\times$150 & 0.7 \\
r' & 09 Nov. 2010 & 5$\times$150 & 0.5 & 05 Jan. 2011 & 4$\times$100 & 0.6 \\
i' & 09 Nov. 2010 & 5$\times$150 & 0.4 & 05 Jan. 2011 & 4$\times$100 & 0.6 \\
z' & 17 Nov. 2010 & 5$\times$300 & 0.6 & 05 Jan. 2011 & 4$\times$300 & 0.6 \\
\hline
\end{tabular}
\end{table*}

We performed spatial dithering between the individual exposures of the fields  
in order to facilitate the removal of cosmic rays and to fill 
in the gaps between the CCD chips. The raw images were processed using the 
GMOS package within IRAF (e.g. \textsf{gprepare}, \textsf{gbias}, 
\textsf{giflat}, \textsf{gireduce} and \textsf{gmosaic}). The appropriate bias 
and flat-field images were obtained from the Gemini Science Archive (GSA) as 
part of the standard GMOS baseline calibrations. The images taken in the filter
$z'$ show significant fringing patterns which were removed by subtracting a 
master fringe frame that was created using \textsf{gifringe} and 
\textsf{girmfringe} tasks. The resulting images corresponding to the same 
filter and the same field, were co-added using the task \textsf{imcoadd} 
in order to obtain the final images used for further analysis. 
The photometric data obtained from them were later calibrated to the
photometric system of the {\it Sloan Digital Sky Survey} (SDSS).

\begin{figure}
\centering
\includegraphics[scale=0.45]{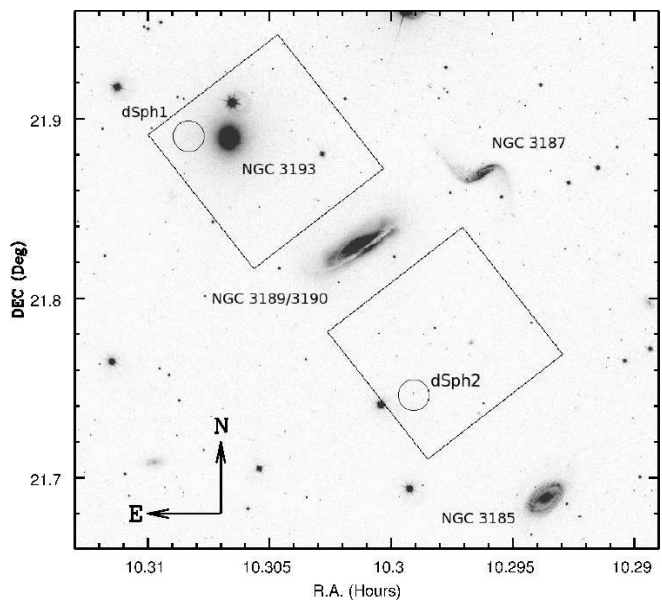}\\
\caption{19.7 $\times$ 18.3 arcmin mosaic in the $g'$ filter from SDSS DR12 
showing the sky region of HCG\,44. The black rectangles depict the 
two Gemini-GMOS frames used in this work ($\sim 5.5$ arcmin on a
side). The black circles indicate
the location of the two dSph candidates. North is up and East to the left.}
\label{SDSS}  
\end{figure}

\begin{figure}
\center
\includegraphics[scale=0.28]{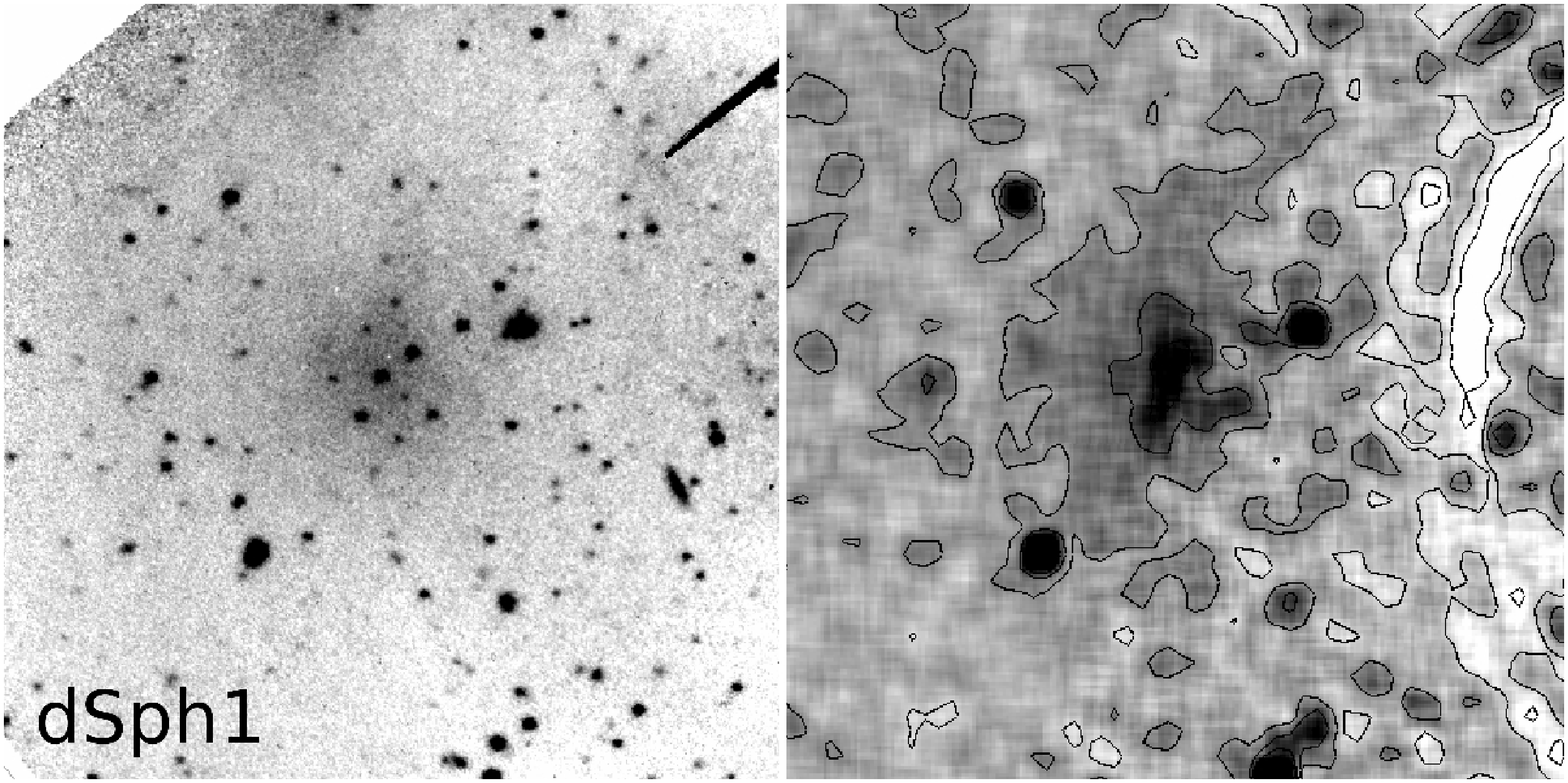}\\
\includegraphics[scale=0.295]{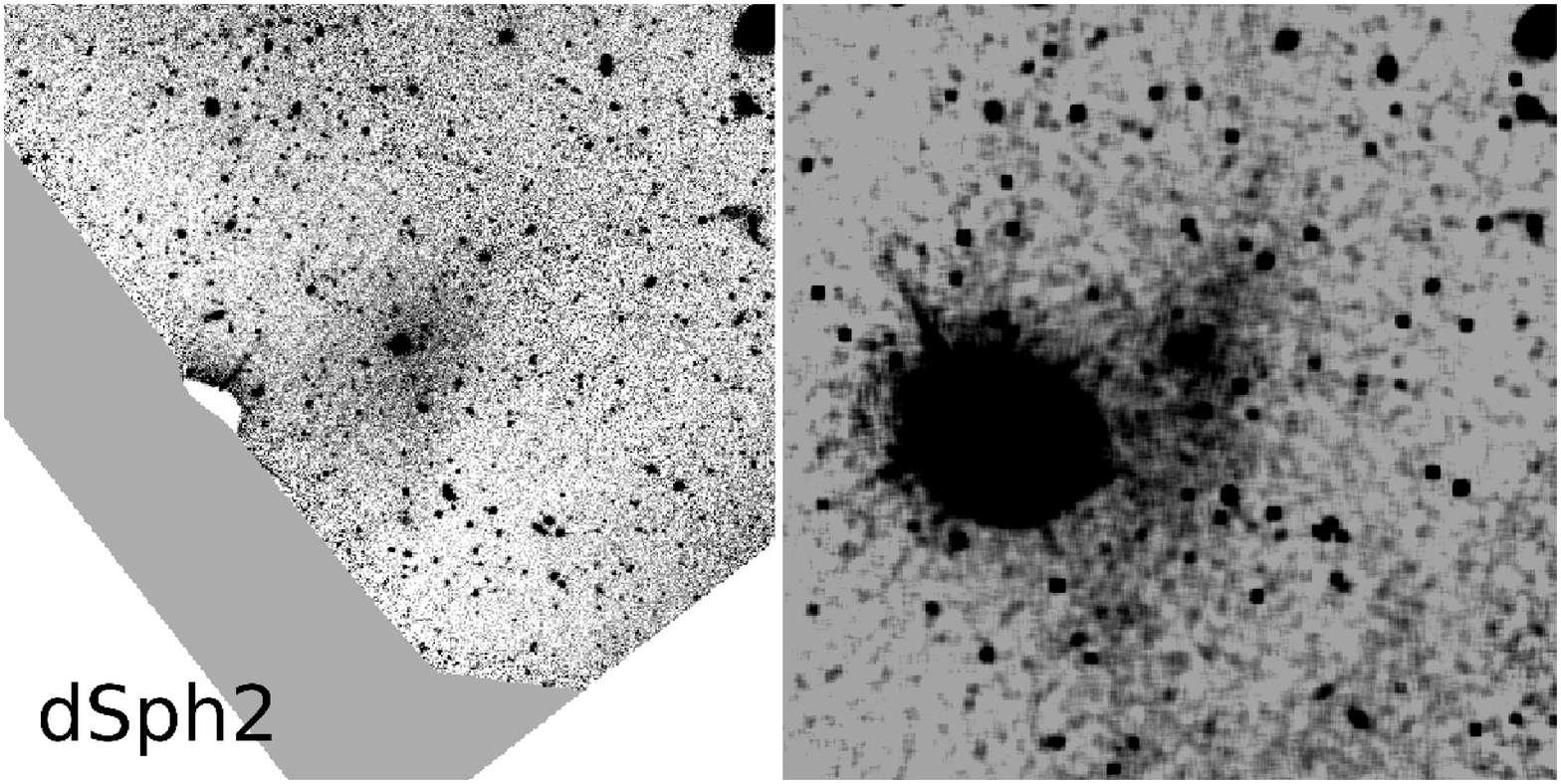}\\
\caption{GEMINI-GMOS {\it (left)} and SDSS DR12 {\it (right)} 
$g'$ images of the LSB galaxies located, in projection, within the halo of NGC\,3193 (top, 
90.3 arcsec on a side; dSph1) and between NGC\,3189/3190 and NGC\,3185 
(bottom, 250 arcsec on a side; dSph2). For clarity, we added level contours 
in the case of the SDSS image of dSph1. The images of SDSS DR12 have been 
processed with the IRAF task \emph{boxcar} considering a box of 
10$\times$10 pixels. As in Figure\,\ref{SDSS}, North is up and East to the left.}
\label{dSphs}
\end{figure}

We detected two LSB structures in our Gemini 
frames, observable in the four filters. With the aim at confirming 
the existence of both LSB 
structures, we downloaded $g',r',i',z'$ images of the same regions of 
the sky from the 
database of the SDSS Data Release 12 (DR12)
\citep{2015ApJS..219...12A}. The SDSS camera consists 
of 30 2048 $\times$ 2048 SITe/Tektronix 49.2 mm square CCDs, which provide a 
FOV of 13.51 $\times$ 8.98 arcminutes and a scale of 0.396 arcsec 
pixel$^{-1}$. The integration time of these images is 54 sec. In 
Figure\,\ref{SDSS} we show the location of the two dSph candidates
in a SDSS mosaic. In Figure\,\ref{dSphs} we show the images of the
low-surface brightness galaxies, both from SDSS and Gemini frames. 
It can be seen that both LSB objects are detectable in both sets
of images.

\section{Results}
\label{resultados}

\subsection{DSph galaxy in the halo of NGC\,3193}
\label{dSph1}

In order to obtain the brightness profiles of both LSB 
galaxies, we worked with ELLIPSE within IRAF on the GEMINI-GMOS 
sky-subtracted images. 
In the particular case of the object
within the halo of NGC\,3193 (hereafter dSph1; top panels in 
Figure\,\ref{dSphs}), we first modeled the bright 
elliptical galaxy, the halo of which covers the whole frame, 
and subtracted it from the original images. In the galaxy-subtracted images, 
we modeled the light of the bright star located next to the elliptical. This
model was later subtracted from the original images, and this star-subtracted
images were used to obtain the final model of the elliptical galaxy. The 
images over which we worked to obtain the profiles of
the LSB object were the star-galaxy-subtracted ones.

To model dSph1, we masked all the objects in the images and we used a 
fixed center, ellipticity and 
position angle, keeping the ellipticity almost circular ($\epsilon=0.05$).
Once we obtained the final table of parameters, we corrected the sky level 
by constructing curves of growth, i.e.
we plotted the integrated flux within the different apertures versus
the semimajor axis of the apertures. The correction to the sky level
of an image is the value for which these curves display an asymptotic flat 
behaviour to infinity. Therefore, we added or subtracted
different constant values (smaller than 2 per cent of the original 
estimated level) to the integrated fluxes until we got asymptotically
flat curves. These corrected fluxes provided the final 
$g'$, $r'$, $i'$ and $z'$ surface brightness profiles
shown in the upper panel of Figure\,\ref{perfiles}. 
They are extinction corrected using the values provided by NED from
the work of \citet{2011ApJ...737..103S}. 
We aware the reader
that the $r'$ and $i'$ frames are affected by strong bleeding patterns due to a 
bright star. Therefore, the brightnesses in these filters should be taken with
caution.

In the case of dSph1, the curves of growth of the different filters 
stabilize at an equivalent radius\footnote{$r_{\rm eq}=sma*\sqrt{(1-\epsilon)}*scale$} 
$r_{\rm eq}\sim26.0$ arcsec. Therefore, we 
will take this radius as the total equivalent radius of the galaxy, and as 
the aperture within which we will measure its total integrated magnitudes.
The surface brightness reached at this aperture is 
$\mu_{\rm g'}\sim30$ mag arcsec$^{-2}$. 

If we assume that this LSB object is located at the distance of NGC\,3193  
($D=33.6$ Mpc, \citealp{2013AJ....146...86T}), 
we have a physical scale of 0.163 kpc arcsec$^{-1}$ and the total
projected radius of this galaxy would be 4.2 kpc. If this galaxy were at 
the distance of HCG\,44 (we will assume a mean distance of $D=23.6$ Mpc, 
$\rm(m-M)=31.86$,
from those of NGC\,3185 and NGC\,3189/3190 shown in Table\,\ref{distancias}) 
the physical scale would be 0.115 kpc arcsec$^{-1}$ and
the radius would translate into 3.0 kpc. 

The projected separation
between dSph1 and NGC\,3193 is 1.4 arcmin (13.7 kpc at the distance
of NGC\,3193). The separation between dSph1 and NGC\,3189/3190 is 
$\sim6.8$ arcmin (46.9 kpc at the distance of HCG\,44).

The brightness profiles of dSph1 are rather noisy and they display 
extremely low signal-to-noise ratio ($S/N=1.4$ per binned pixel
in the inner zone of the $g'$ profile). Therefore, instead of performing
a numerical integration of the brightness profile, we will take as an 
approach to the total integrated magnitudes those obtained from the 
total flux of the stabilized curve of growth. 
These magnitudes are listed in Table\,\ref{parameters}. 
To estimate the errors to the magnitudes, we followed 
the procedure used by \citet{1994ApJS...93..397C} and 
\citet{1999A&A...345..403C} in their studies
of LSB galaxies in the Fornax cluster and NGC\,5044 group. We evaluated
the variations of the background near the galaxy by fitting
a surface to the mean sky value. In the case of dSph1, we obtained that these 
variations are $\sim 0.2\%$ of that value. As this value resulted similar to 
that of the mean error in the sky, we changed 
the curves of growth by adding and sustracting this amount of sky value, and 
calculated the upper and lower limits of the total magnitudes from these new 
curves of growth.

Considering the effective radius as the equivalent radius that contains
half of the total flux of the galaxy, we obtain $r_{\rm eff}\sim11.5$ arcsec.
Depending on the distance considered to the galaxy, this translates into 
$r_{\rm eff}\sim 1.9$ kpc (NGC\,3193) or $r_{\rm eff}\sim1.3$ kpc (HCG\,44).
According to Figure\,\ref{mueff_dSphs}, if dSph1 were indeed at the distance
of NGC\,3193 or HCG\,44, its $r_{\rm eff}$ would 
be in agreement to those of the "ultra-diffusse galaxies" (UDGs) reported by 
\citet{2015ApJ...798L..45V} in the Coma cluster, and it would be quite large in
comparison to those reported for most of the LG dSphs (\citealp{2012AJ....144....4M}
and references therein). Figure\,\ref{CMR_dSphs}, on
the other hand, show that this object, shifted to the Virgo cluster distance, 
would not follow the red-sequence (RS) of Virgo's early-type galaxies, in contrast with 
the UDGs reported by \citet{2015ApJ...807L...2K} in Coma which do follow Comas' 
RS. 

We built a model of the galaxy through the IRAF task {\it bmodel}.
Its subtraction from the galaxy-star subtracted images left no residuals,
which indicates the abscence of star-forming regions, internal substructures
or distorted isophotes, which would be evidence of interaction.

\subsection{DSph galaxy between NGC\,3189/3190 and NGC\,3185}

After performing several tests, and due to the extremely
low surface brightness of the galaxy (S/N=0.5 per binned 
pixel in the inner zone of 
the $g'$ profile), we considered fixed circular isophotes to 
obtain the brightness profile of the dSph galaxy located 
within the HCG\,44 region (hereafter dSph2; lower panels of 
Figure\,\ref{dSphs}). 
As in the case of dSph1, 
we worked on sky-subtracted images in which we masked all the extended
and unresolved objects, 
and we built the curve of growth in order to correct the initially
assumed sky value. In this case, the curve of growth in the $g'$
filter stabilizes at
$r_{\rm eq}\sim54$ arcsec ($\mu_{\rm g'}\sim30$ mag arcsec$^{-2}$). 
If the galaxy were located at the 
distance of the HCG\,44, this total
equivalent radius would translate into $r=6.2$ kpc. 
In that case, this would imply that
this object would be a large LSB galaxy. 

\begin{figure}[h!]
\centering
\includegraphics[scale=0.45]{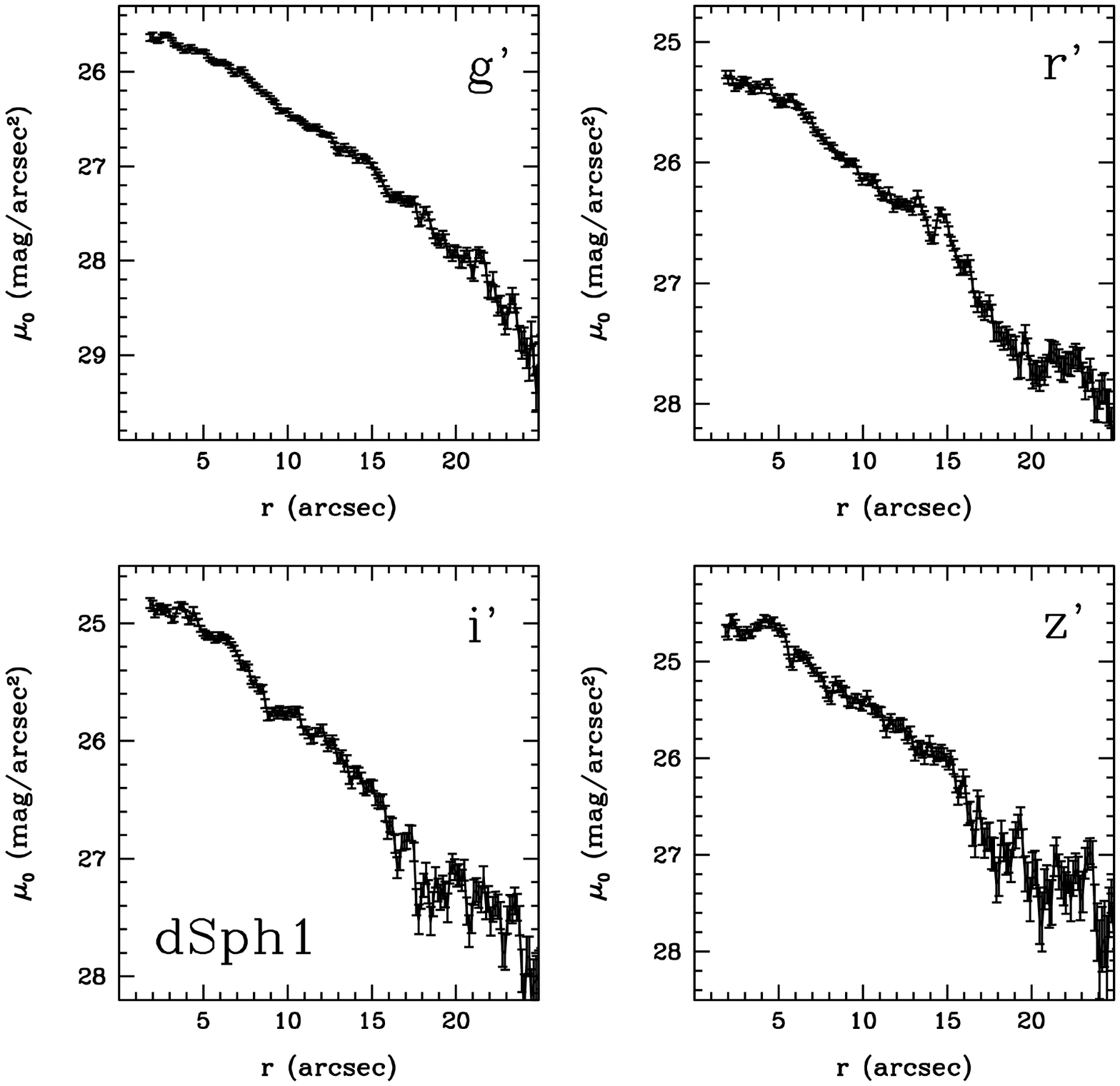}
\includegraphics[scale=0.45]{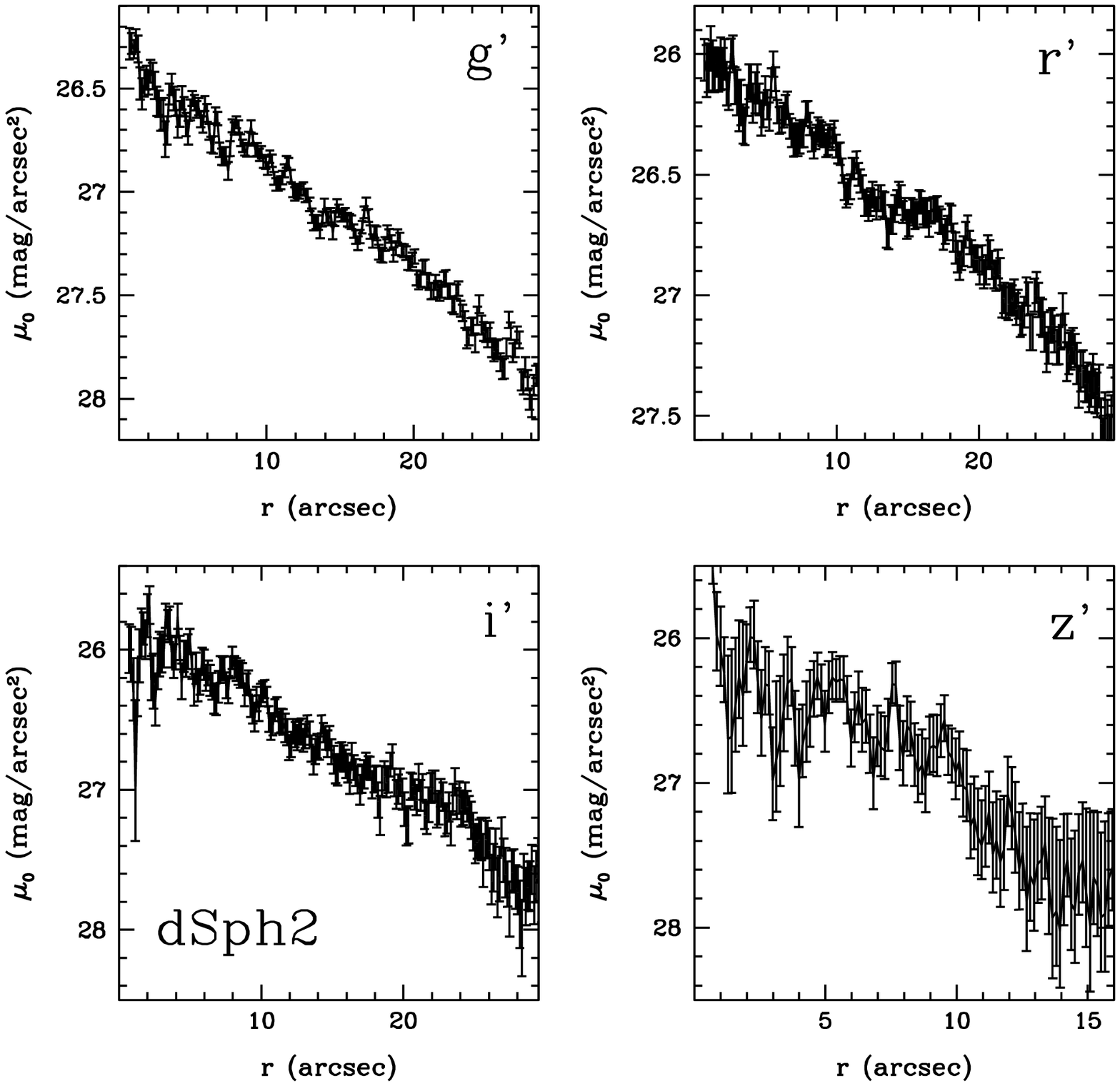}
\caption{Extinction corrected $g'$, $r'$, $i'$ and $z'$ surface brightness 
profiles of dSph1 {\it (top)} and dSph2 {\it (bottom)}.}
\label{perfiles}  
\end{figure}

We show the extinction corrected brightness profiles in the lower panel
of Figure\,\ref{perfiles}. In this case, as the galaxy is extremely faint, 
the surface brightness profiles are rather noisy. The integrated 
magnitudes, obtained from the total flux of the stabilized curve of growth,
must be taken with causion and are presented in 
Table\,\ref{parameters}. Magnitude errors were obtained in a similar
manner to those of the integrated magnitudes of dSph1 (see Section\,\ref{dSph1}). 

\begin{figure*}[h!]
\centering
\includegraphics[scale=0.9]{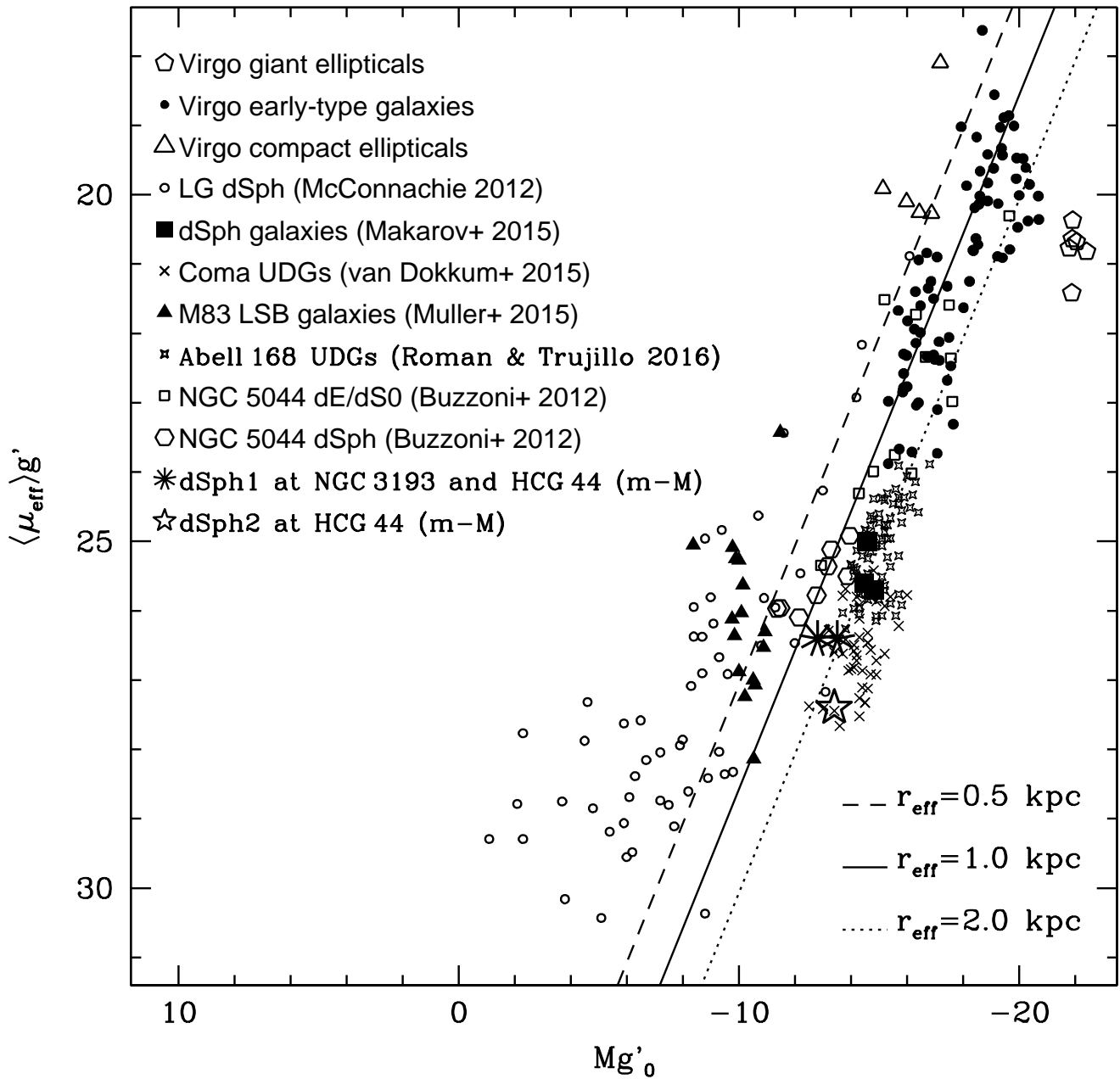}\\
\caption{$\langle\mu_{\rm eff}\rangle$-luminosity diagram of the early-type galaxies
in the central region of the Virgo cluster \citep{2013ApJ...772...68S},
showing the location of the LSB galaxies presented
in this work (assuming they are at the distances of NGC\,3193 and/or HCG\,44), 
plus different samples of dSph and ultra-diffuse galaxies
reported in the literature. In the case of dSph1, we plot
two locations considering the distance modulii of NGC\,3193 and 
HCG\,44.}
\label{mueff_dSphs}  
\end{figure*}

\begin{figure}[h!]
\centering
\includegraphics[scale=0.45]{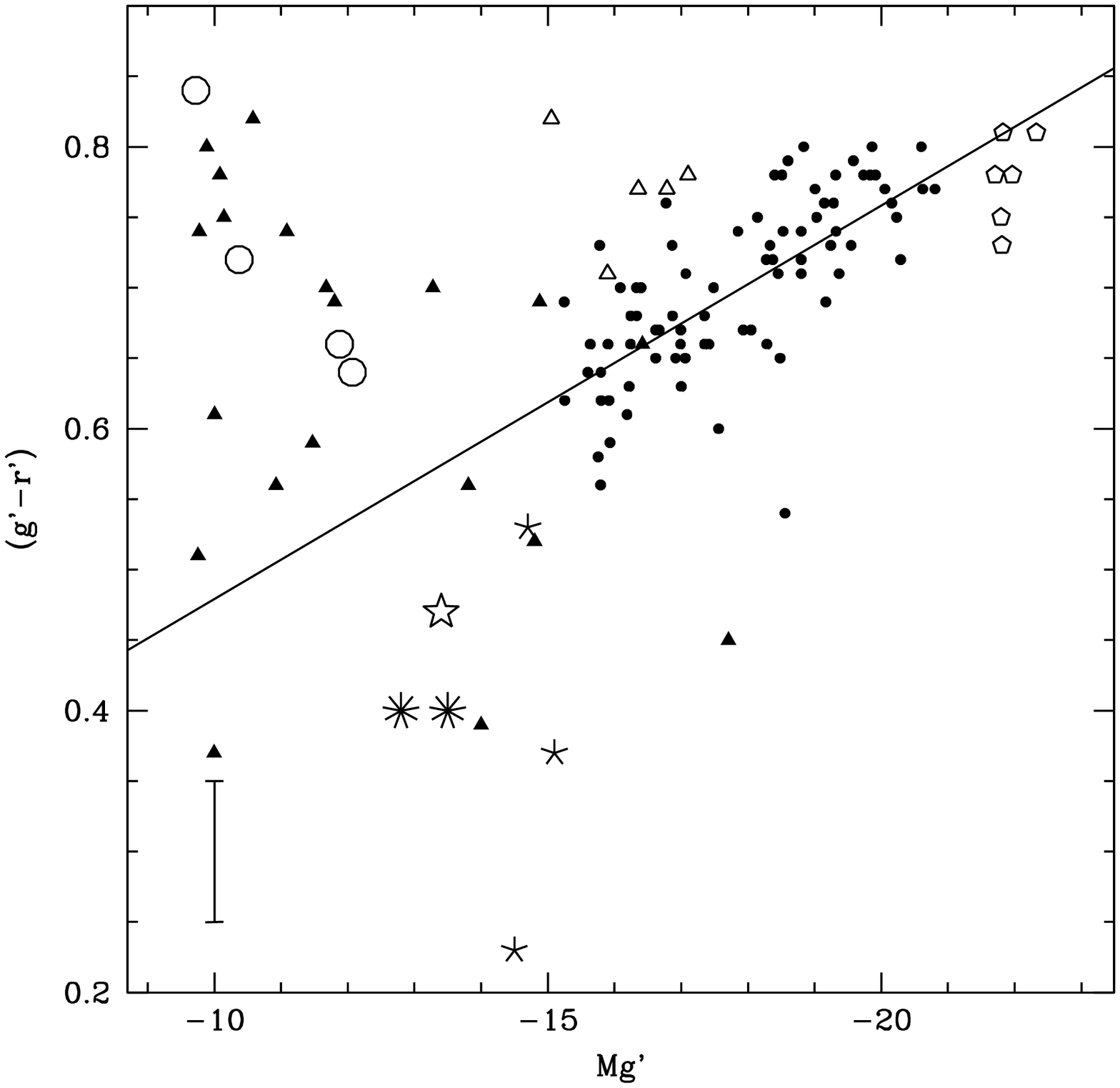}\\
\caption{Color-magnitude diagram of the early-type galaxies in 
the central region of the Virgo cluster \citep{2013ApJ...772...68S}, showing
the location of the dSph galaxies presented in this work {\it (asteriscs and open star)} 
assuming they are at the distances of NGC\,3193 and/or HCG\,44. We also include  
the galaxies reported by 
\citet{2015A&A...581A..82M} {\it (filled stars)}, \citet{2015A&A...583A..79M}
{\it (filled triangles)} and \citet{2012AJ....144..190L} {\it (big open circles)}. 
The symbols code for Virgo galaxies is the same as in Figure\,\ref{mueff_dSphs}. The
big vertical errorbar represents the mean error in $(g'-r')$ (0.05 mag) for all 
galaxies included in the plot, excepting the LSB galaxies reported in this work. 
}
\label{CMR_dSphs}  
\end{figure}

The effective radius of dSph2 in the $g'$ filter is $r_{\rm eff}\sim19$ arcsec.
If dSph2 were indeed at the distance of HCG\,44, it would translate 
into $r_{\rm eff}\sim2.2$ kpc. 
Similarly to dSph1, the $r_{\rm eff}$ of dSph2 would be in agreement to those
reported for Coma's UDGs and would be significantly
larger than those of LG dSphs (Figure\,\ref{mueff_dSphs}), if it were a member
of HCG\,44. In addition, it would not follow Virgo's RS. 

The projected separation between dSph2 and NGC\,3185 is 5.5 arcmin, while
the separation between dSph2 and NGC\,3189/3190 is 5.6 arcmin. At the 
distance of the HCG\,44, these separations would translate into 37.9 kpc
and 38.6 kpc, respectively.

\begin{figure}[h!]
\centering
\includegraphics[scale=0.45]{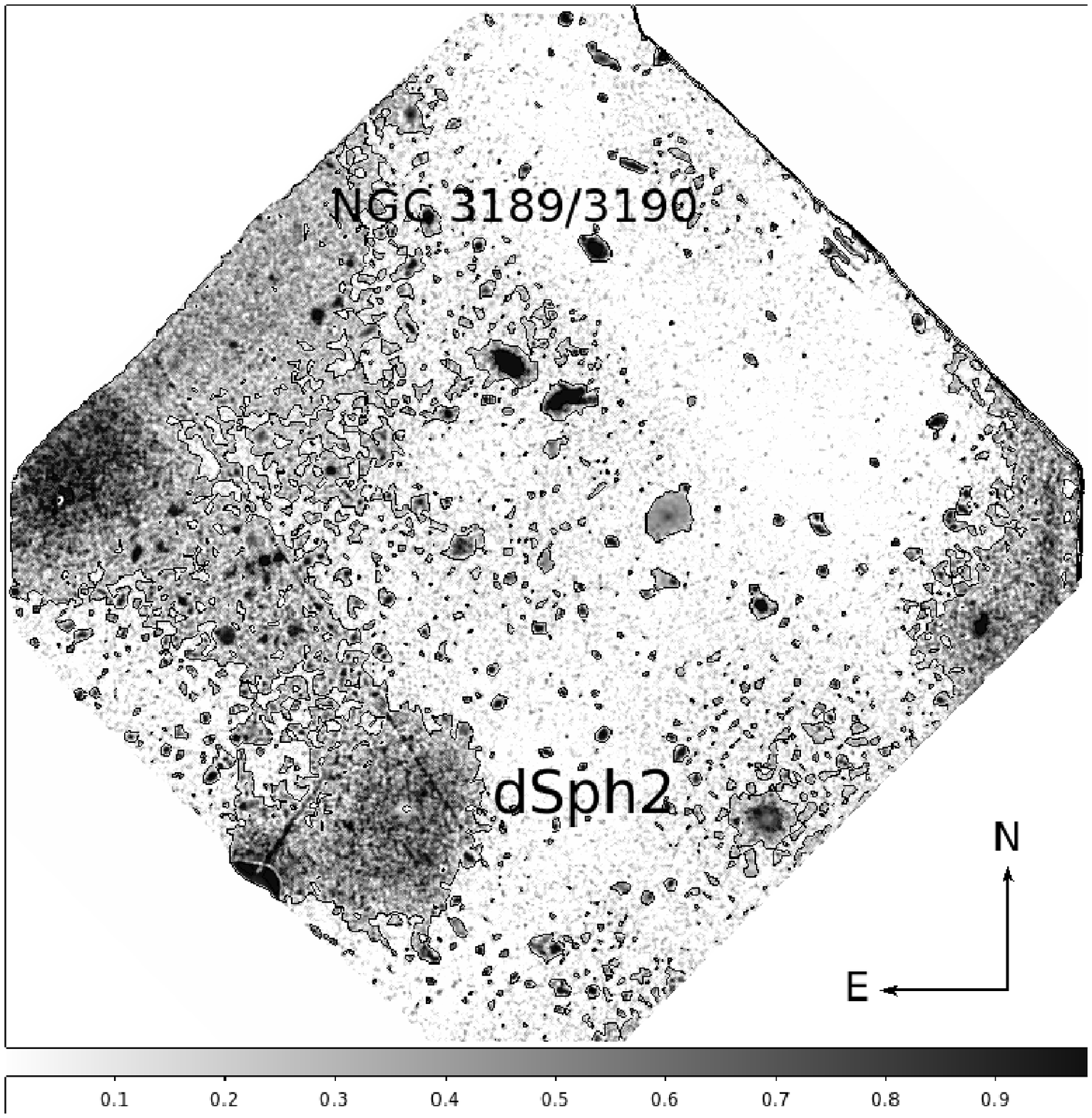}\\
\caption{$(g'-r')$ color map of the GEMINI-GMOS field (5.5$\times$5.5 arcmin) 
containing dSph2. We can see
that there seems to be a LSB structure linking dSph2 to the halo
of NGC\,3189/3190. This structure displays a similar color to that of the 
external regions of dSph2 ($(g'-r')\sim0.33$). This might be evidence of a tidal
origin for dSph2. As in Figure\,\ref{SDSS}, North is up and East to the left.}
\label{dSph2+3189}  
\end{figure}

The sustraction of a model of the galaxy from the sky-subtracted images, 
left no residuals which
is evidence of the abscence of star-formation regions, inner structures
or distorted isophotes.
However, as it is shown in Figure\,\ref{dSph2+3189}, we detected in
the $(g'-r')$ color map of the GEMINI-GMOS field containing dSph2,
what seems to be a LSB structure linking dSph2 to the halo of 
NGC\,3189/3190. This structure displays $(g'-r')\sim0.33$, a similar color to 
that of the external regions of dSph2. This color is much bluer than those expected
if this structure were due to, for example, MW cirrus ($(g'-r')=1.33-2.03$ mag, 
\citealp{2012AJ....144..190L}).

\begin{table}
\caption{Basic information of the LSB objects analyzed in this paper. 
Aparent magnitudes and colors are not extinction or reddening corrected.}
\label{parameters}
\centering
\begin{tabular}{@{}lll@{}}
\hline\hline
  & \multicolumn{1}{c}{dShp1} & \multicolumn{1}{c}{dSph2} \\
\hline
R.A. (J2000) & 10:18:31 & 10:17:58 \\
DEC (J2000)  &  21:53:41 & 21:44:53 \\
$r_{\rm tot}$ (arcsec) &  26.0 &  54.0 \\
$r_{\rm eff}$ (arcsec) &  11.5 &  24.0 \\
$m_{\rm g'}$ (mag) &  19.1$\pm$0.4 & 18.5$\pm$0.2\\
$m_{\rm r'}$ (mag) &  18.7$\pm$0.5 & 18.1$\pm$0.2\\
$m_{\rm i'}$ (mag) &  18.4$\pm$0.3 & 18.2$\pm$0.4\\
$m_{\rm z'}$ (mag) &  17.9$\pm$0.5 & 19.7$\pm$1.1\\
$(g'-r')$ (mag) &   0.4$\pm$0.6 &  0.4$\pm$0.3 \\
$\langle \mu_{\rm eff_{\rm g'}}\rangle$ (mag arcsec$^{-2}$) & 26.4  & 27.4 \\
\hline
\end{tabular}
\tablefoot{\\
(1): $r_{\rm tot}$ is the equivalent radius at which the curve of growth in 
the $g'$ filter stabilizes.\\
(2): $r_{\rm eff}$ is the equivalent radius that contains half of the flux within $r_{\rm tot}$\\
(3): Total integrated magnitudes are taken where the curves of growth stabilize,
 which might happens at different radius in different filters.\\
(4): $(g'-r')$ colors are taken at the total radius of the galaxies as $g'$ and
$r'$ curves of growth stabilize at this radius.\\
(5):$\langle\mu_{\rm eff}\rangle=mag+2.5~log(2 \pi r_{\rm eff}^2)$.
}
\end{table}
 
\section{Globular Clusters?}
\label{Globulars}

Globular clusters (GCs) are stellar systems believed to be formed
during the first stages of the formation of their host galaxies. 
They are considered as good examples of single stellar populations
and, therefore, they are expected to provide information about the 
processes experienced by the galaxies in which they reside.

Our images have been obtained with the intention of studying the
population of GCs within the HCG\,44 (Faifer et al., in preparation). 
With this in mind, we searched for GC candidates in the neighbourhoods
of dSph1 and dSph2. To this aim, we performed point spread function
(PSF) photometry in both fields. As a
next step, we selected GC candidates according to the stellarity
index provided by SExtractor \citep{1996A&AS..117..393B}, taking
0.5 as the value that separates resolved from unresolved
sources. Finally, from colors presented by \citet{2013MNRAS.431.1405F},
we adopted the following ranges to select GC candidates:

\begin{itemize}
\item[]$0.0<(r'-i')<0.5$
\item[]$0.4<(g'-r')<0.9$
\item[]$0.5<(g'-i')<1.5$
\item[]$0.7<(g'-z')<1.7$
\item[]$0.1<(r'-z')<1.0$
\item[]$21.0<i'< 24.0$.
\end{itemize}

The upper brightness limit in the $i'$ filter is taken considering that the 
completness experiments show that the sample is complete for $i'<24$. The 
lower limit implies that we are including as GC candidates all objects with 
$M_{i'}>-12$ mag \citep{2011MNRAS.416..155F}. It is worth mentioning
that selecting GC candidates from high-quality images (FWHM < 1 arcsec) 
by using different color indeces has provided very clean samples as it is 
shown by the spectroscopic studies by \citet{2006MNRAS.373..157B} 
and \citet{2006MNRAS.366.1253P}.

\subsection{GC candidates around dSph1}

As it was mentioned in the previous sections, dSph1 is located, in projection,
within the halo of the bright elliptical galaxy NGC\,3193, which shows a rich GC
system. dSph1 is located at a projected distance of 1.4 arcmin from the elliptical.
Therefore, it is
hard to certainly identify a GC population linked to this LSB object. 
The mean density of GC candidates at that galactocentric
distance is 14 objects arcmin$^{-2}$. In the left panel of 
Figure\,\ref{dSphs_GCs} we show 
a Gemini-GMOS image of 2.0 $\times$ 2.0 arcmin centered on dSph1,  
where we depict with circles the GC 
candidates identified in the zone.

We are not able to detect an overdensity around dSph1, but this is expected 
from the low number of GCs reported for low-luminosity galaxies 
\citep{2010MNRAS.406.1967G}. Therefore, as a tentative approach, we defined 
an area where we expect to find some genuine GC of dSph1. To this aim, we selected 
the area inside the circle of radii $r_{\rm tot}$ (see Table\,\ref{parameters}). 
There are three candidates which fill the colour and magnitude selection previously mentioned
(black circles). Besides, there are four unresolved and quite bright objects with $g'=23-23.8$ mag 
which appears overimpossed to dSph1.
Unfortunately, we are 
not able to obtain $(g'-r')$ and $(g'-i')$ colors for two of them as a consequence of 
a strong bleeding pattern in the $r'$ and $i'$ frames 
due to a bright star present in the field (white circles in Figure\,\ref{dSphs_GCs}). 
However, for two of them, it was possible to obtain $(g'-z')$ colors 
(small black circles in Figure\,\ref{dSphs_GCs}).

In Table\,\ref{GC_dSph1} we show
the basic information extracted for the unresolved objects detected in the
region occupied by dSph1. We can see that all of the sources with measured
colors, show typical values of metal-poor GCs (i.e. blue GCs). It is worth
mentioning that GCs confirmed in low-mass galaxies
are mainly blue globulars \citep{2005A&A...442...85S}. Furthermore, 
the objects observed overimposed 
to dSph1 would be rather bright sources ($M_V<-9$ mag) at the distances of 
NGC\,3193 or HCG\,44. Therefore, if dSph1 were located at these distances, 
these unresolved objects might be nuclear massive clusters similar to those reported in 
other dSph galaxies (e.g. \citealp{2008ApJ...674..909P,2009MNRAS.392..879G}).

\begin{table*}
\caption{Basic information of the unresolved sources detected in the sky region
occupied by dSph1.}
\label{GC_dSph1}
\centering
\begin{tabular}{ccccccc}
\hline\hline
& \multicolumn{1}{c}{R.A.} & \multicolumn{1}{c}{DEC} & \multicolumn{1}{c}{$g'$} & \multicolumn{1}{c}{$(g'-r')$} & \multicolumn{1}{c}{$(g'-i')$} & \multicolumn{1}{c}{$(g'-z')$} \\
& \multicolumn{1}{c}{(J2000)} & \multicolumn{1}{c}{(J2000)} & \multicolumn{1}{c}{(mag)} & \multicolumn{1}{c}{(mag)} & \multicolumn{1}{c}{(mag)} & \multicolumn{1}{c}{(mag)} \\
\hline
dSph1\_1  &   10:18:29.874  & 21:53:51.45 &  23.742$\pm$0.016 &             -  &            -  &     -  \\
dSph1\_2  &   10:18:30.184  & 21:53:46.90 &  23.253$\pm$0.012 &             -  &            -  &     -  \\
dSph1\_3  &   10:18:30.603  & 21:53:43.73 &  23.146$\pm$0.011 &             -  &            -  & 0.98$\pm$0.02 \\   
dSph1\_4  &   10:18:30.864  & 21:53:40.95 &  23.526$\pm$0.014 &             -  &            -  & 0.83$\pm$0.03 \\
dSph1\_5  &   10:18:29.236  & 21:53:46.98 &  24.726$\pm$0.029 &  0.58$\pm$0.03 & 0.80$\pm$0.03 & 1.05$\pm$0.06 \\
dSph1\_6  &   10:18:29.772  & 21:53:35.16 &  24.069$\pm$0.014 &  0.60$\pm$0.02 & 0.92$\pm$0.02 & 1.19$\pm$0.03 \\
dSph1\_7  &   10:18:29.954  & 21:53:21.78 &  24.122$\pm$0.017 &  0.58$\pm$0.02 & 0.88$\pm$0.02 & 1.16$\pm$0.03 \\
\hline
\end{tabular}
\end{table*}

\subsection{GC candidates around dSph2}

\begin{figure*}
\center
\includegraphics[scale=0.74]{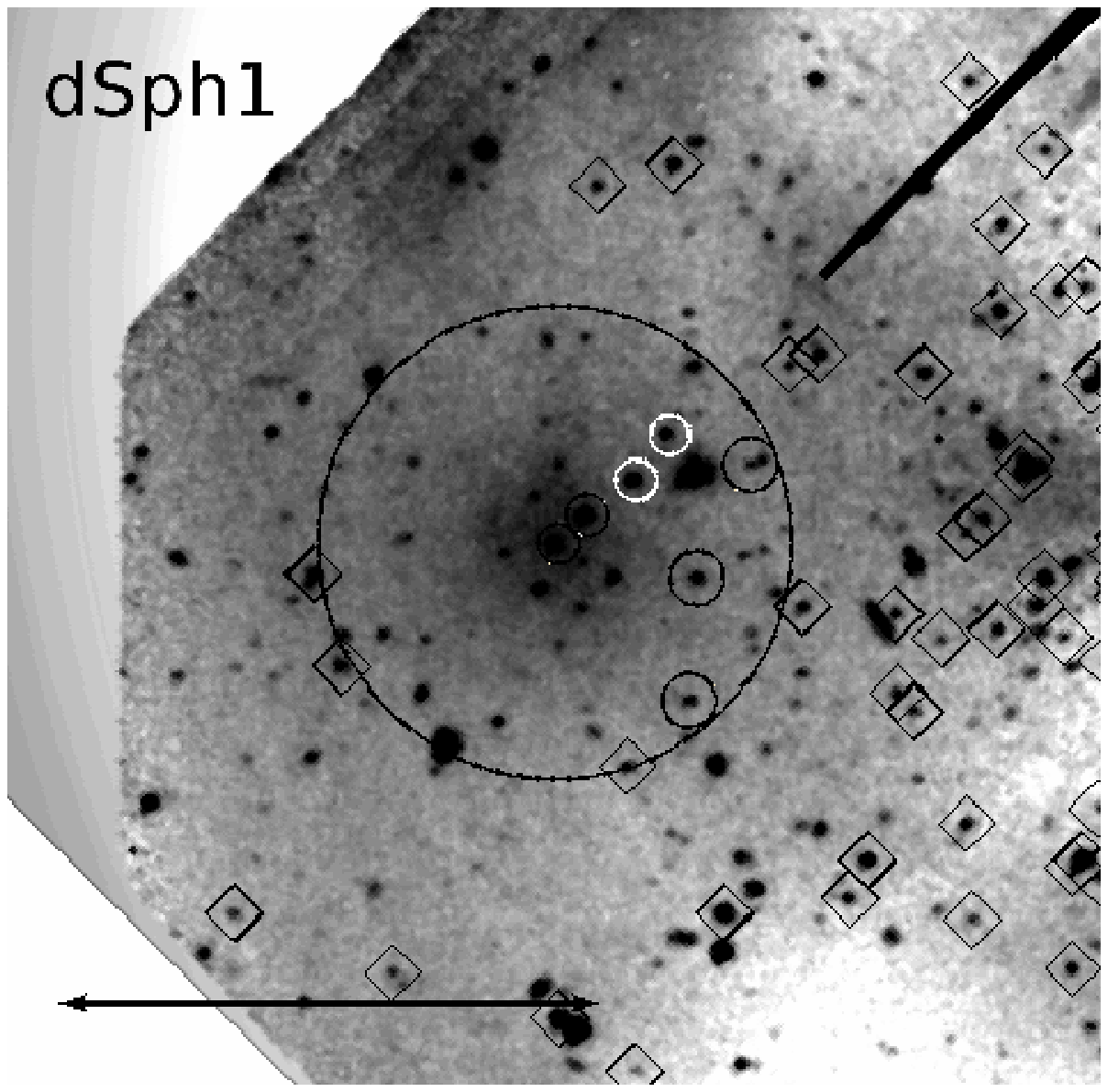}
\includegraphics[scale=0.74]{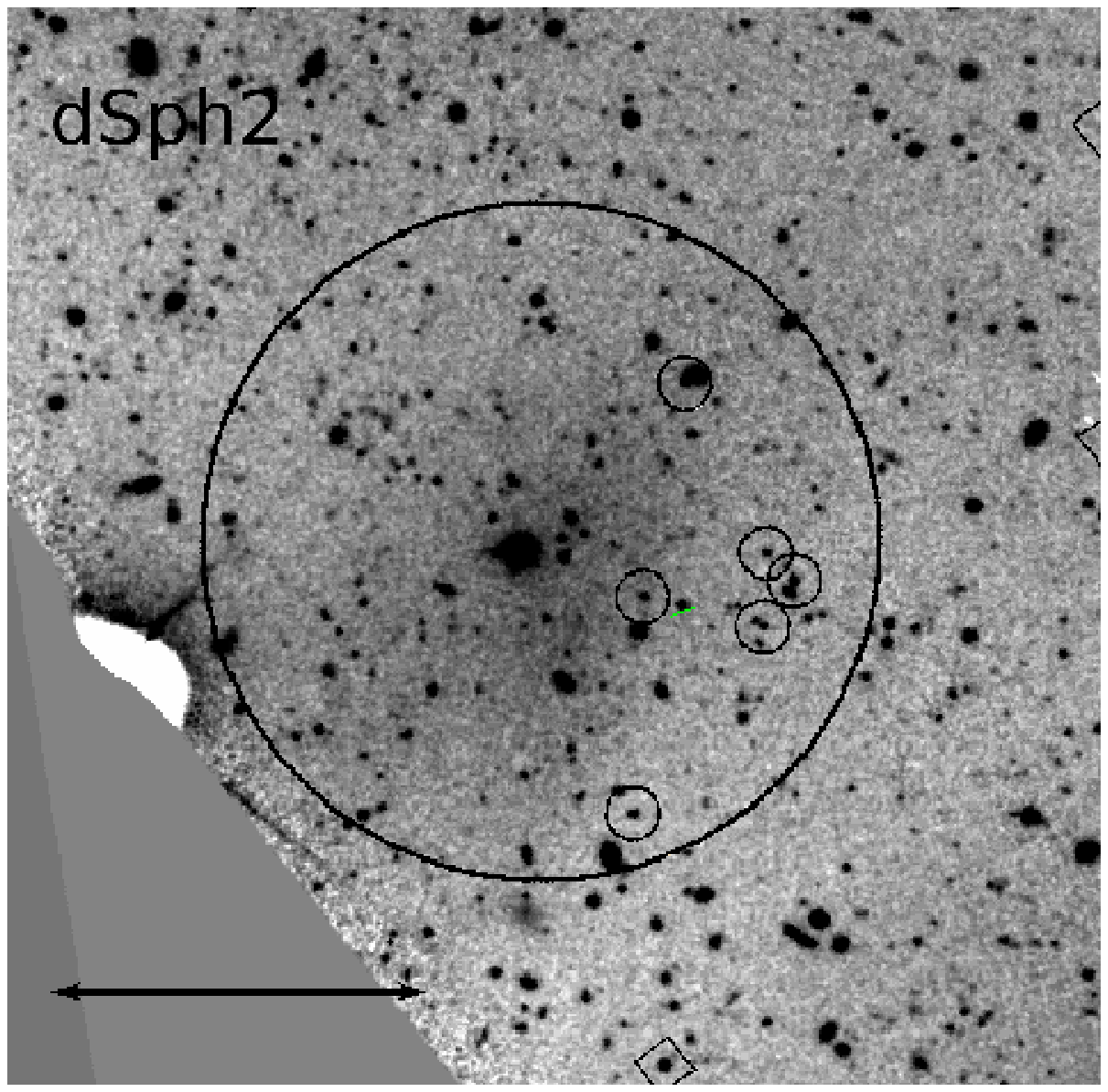}
\caption{The same GEMINI-GMOS $g'$ images shown in Figure\,\ref{dSphs},
but now showing the location of the GC candidates within the fields (rhomboids),
and the GC candidates located within the total radii of the dSph galaxies (small circles). The big
circles in both images depict the total radii measured for the galaxies.
The white circles in the field of dSph1 are GC candidates with no (g'-r') and (g'-i') color 
information due to a strong bleeding pattern caused by a near bright star. The horizontal
bars depict a scale of 1 arcmin in each frame.  
As in Figure\,\ref{SDSS}, North is up and East to the left.}
\label{dSphs_GCs}
\end{figure*}

DSph2 is more distant from the bright galaxies present in the region
of HCG\,44 than dSph1. Therefore, the object density in the zone
is lower (1.08 object arcmin$^{-2}$) than that of the neighbourhoods of dSph1. 
In the right panel of Figure\,\ref{dSphs_GCs} we can see that, 
within a radius equal to the equivalent
radius of dSph2, there are six unresolved sources with colours similar to 
those of GCs. As in the case of dSph1, it cannot be detected any overdensity
associated with the galaxy.

As shown in Table\,\ref{GC_dSph2}, all the detected objects would be blue GCs,
similar to those found in the frames of dSph1. From a nearby control field, 
we have estimated that the expected contamination in the zone of HCG\,44
is 0.3 object arcmin$^{-2}$. 
These blue GC candidates might belong to NGC\,3189/3190,
or to dSph2 if this galaxy were located at the distance of HCG\,44. In the latter case,
they would show
an assymetric distribution, similar to those found in other dSph galaxies
(e.g. \citealp{2015A&A...581A..84T,2016ApJ...819L..20B}).

\begin{table*}
\caption{Basic information of the unresolved sources detected in the sky region
occupied by dSph2.}
\label{GC_dSph2}
\centering
\begin{tabular}{ccccccc}
\hline\hline
& \multicolumn{1}{c}{R.A.} & \multicolumn{1}{c}{DEC} & \multicolumn{1}{c}{$g'$} & \multicolumn{1}{c}{$(g'-r')$} & \multicolumn{1}{c}{$(g'-i')$} & \multicolumn{1}{c}{$(g'-z')$} \\
& \multicolumn{1}{c}{(J2000)} & \multicolumn{1}{c}{(J2000)} & \multicolumn{1}{c}{(mag)} & \multicolumn{1}{c}{(mag)} & \multicolumn{1}{c}{(mag)} & \multicolumn{1}{c}{(mag)} \\
\hline
dSph2\_1  & 10:17:56.033   & 21:45:22.65 & 24.335$\pm$0.029 & 0.63$\pm$0.04 & 0.81$\pm$0.05 & 1.07$\pm$0.04 \\
dSph2\_2  & 10:17:56.471   & 21:44:47.96 & 23.997$\pm$0.015 & 0.61$\pm$0.02 & 0.84$\pm$0.02 & 0.92$\pm$0.02 \\
dSph2\_3  & 10:17:55.059   & 21:44:55.35 & 24.292$\pm$0.016 & 0.51$\pm$0.02 & 0.76$\pm$0.02 & 0.85$\pm$0.03 \\   
dSph2\_4  & 10:17:54.737   & 21:44:50.96 & 23.558$\pm$0.015 & 0.62$\pm$0.02 & 0.78$\pm$0.02 & 0.94$\pm$0.02 \\
dSph2\_5  & 10:17:55:082   & 21:44:43.53 & 24.494$\pm$0.021 & 0.58$\pm$0.03 & 0.84$\pm$0.03 & 0.98$\pm$0.04 \\
dSph2\_6  & 10:17:56.548   & 21:44:13.19 & 24.006$\pm$0.016 & 0.54$\pm$0.02 & 0.67$\pm$0.03 & 0.84$\pm$0.03 \\
\hline
\end{tabular}
\end{table*}

\section{Discussion \& Conclusions}
\label{conclusions}

From deep GEMINI-GMOS images, we have detected two extended LSB 
objects in the region of the HCG\,44. 
From our photometric analysis we have found the following results:
\begin{itemize}

\item Both galaxies display smooth morfologies. The sustraction of models
 of the galaxies built from elliptical isophotes with fixed parameters,
left no residuals. This might be evidence 
of the absence of inner structures, star-forming regions and
interactions with the brighter galaxies.\\

\item The colors of these objects are in agreement with those reported
for the dSphs studied by \citet{2015A&A...581A..82M}. However, they are much 
bluer than those of the LSB galaxies detected in M\,83 
\citep{2015A&A...583A..79M} and those of the dSph galaxies analyzed by 
\citet{2012AJ....144..190L} around NGC\,7331. \\

\item If both galaxies were placed at the distances of NGC\,3193 and HCG\,44,
the effective radius and luminosities of both objects would be similar to
those of the ultra-diffuse galaxies (UDGs) reported in different environments. 
These sizes would be significantly larger than those of the dSph galaxies of the LG 
\citep{2012AJ....144....4M} and of the LSB galaxies 
recently identified in the M\,83 subgroup in Centaurus \citep{2015A&A...583A..79M}. \\

\item If dSph1 were at the distance of NGC\,3193, it would be one of the most 
extended LSB objects reported up to date.\\

\item If the LSB structure that seems to link dSph2 with 
NGC\,3189/3190 is real, it would be evidence of a tidal origin for dSph2. \\

\item We have detected several unresolved sources in the frames of 
dSph1 and dSph2 consistent with being blue GCs, but we have not detected any 
overdensity that could be associated to any of the two galaxies. 
However, if dSph1 and dSph2 were low-mass galaxies at the distances of 
NGC\,3193 and/or HCG\,44, it would not be expected a high 
number of GCs belonging to them. If the unresolved objects were indeed GCs, 
and dSph1 and dSph2 were at such distances, dSph1 might contain massive nuclear star 
clusters like those reported by \citet{2009MNRAS.392..879G} in several dSph galaxies, 
and the distribution of the GCs of dSph2 would be assymetric like that shown by the IKN 
dwarf in the M\,81 group \citep{2015A&A...581A..84T}.\\

\item As we are not able to establish the distances to dSph1 and dSph2, it could not
be discarded that these objects were isolated LSB galaxies.

\end{itemize}

According to the {\it Dual Dwarf Galaxy Theorem} postulated by 
\citet{2012PASA...29..395K}, two types of dwarf galaxies exist:
{\it primordial dwarf galaxies}, formed within low-mass dark matter halos,
that, as a consequence, are DM dominated; and 
{\it tidal/ram-pressure dwarfs}, formed in the encounters of galaxies, 
which are devoided of DM. An interesting general result obtained from 
Figure\,\ref{mueff_dSphs}, is that LSB galaxies display a quite
extended range of effective radius, in comparison with {\it ``normal''}
early-type galaxies which show $r_{\rm eff}\sim1.0$ kpc with a low dispersion.
This different behaviour might be pointing to different formation scenarios for
both types of galaxies.

In the context of the discussion about UDGs being failed luminous galaxies or genuine
dwarfs, \citep{2015ApJ...798L..45V,2016arXiv160408024B}, 
if the unresolved objects detected in the frames of dSph1 and dSph2 are indeed GCs
belonging to these galaxies, both galaxies would be in agreement with the scenario of 
\citet{2016arXiv160408024B}.
That is, they would be quenched dwarfs, as all detected unresolved sources
would be blue GCs, and failed luminous galaxies are expected to present both
red and blue GCs and a more rich GC system. 

If both dSph galaxies belong to HCG\,44, and considering
HCG\,44 as composed by four galaxies (NGC\,3185, NGC\,3187, NGC\,3189/90 and
[WMv91]\,1015+2203), then both LSB galaxies would be located in a high-density 
environment, similar to that of the center of rich galaxy clusters.
In addition, they would constitute the $\sim33$\% of the 
total galaxy population of the group, and the $\sim66$\% of the dwarf
galaxy population. These fractions would be in agreement with the estimations
made by \citet{1996ApJ...462...50H} about the amount of dwarf galaxies
of tidal origin expected in Hickson Compact Groups.

We hope that the obtention of spectroscopic data of both dSph galaxies
and their GC candidates in the near future, help to disentangle the 
real nature of all these objects.

\begin{acknowledgements}
We thank the anonymous referee for her/his comments and sugestions which 
helped to improve the content of the paper. We also thank Sergio Cellone
for reading the manuscript and providing interesting discussion.\\

Based on observations obtained at the Gemini Observatory (Program GN-2010B-Q-29), 
which is operated by the Association of Universities for Research in Astronomy, 
Inc., under a cooperative agreement with the NSF on behalf of the Gemini 
partnership: the National Science Foundation 
(United States), the National Research Council (Canada), CONICYT (Chile), the 
Australian Research Council (Australia), Minist\'erio da Ci\^encia, Tecnologia e 
Inova\c c\~ao (Brazil) and Ministerio de Ciencia, Tecnolog\'ia e Innovaci\'on Productiva 
(Argentina).\\

This research has made use of the NASA/IPAC Extragalactic Database (NED) which is operated by 
the Jet Propulsion Laboratory, California Institute of Technology, under contract with the National 
Aeronautics and Space Administration.\\

Funding for SDSS-III has been provided by the Alfred P. Sloan Foundation, the Participating 
Institutions, the National Science Foundation, and the U.S. Department of Energy Office of Science. 
The SDSS-III web site is http://www.sdss3.org/.\\

SDSS-III is managed by the Astrophysical Research Consortium for the Participating Institutions 
of the SDSS-III Collaboration including the University of Arizona, the Brazilian Participation 
Group, Brookhaven National Laboratory, Carnegie Mellon University, University of Florida, the French 
Participation Group, the German Participation Group, Harvard University, the Instituto de Astrofisica de 
Canarias, the Michigan State/Notre Dame/JINA Participation Group, Johns Hopkins University, Lawrence 
Berkeley National Laboratory, Max Planck Institute for Astrophysics, Max Planck Institute for 
Extraterrestrial Physics, New Mexico State University, New York University, Ohio State University, 
Pennsylvania State University, University of Portsmouth, Princeton University, the Spanish Participation 
Group, University of Tokyo, University of Utah, Vanderbilt University, University of Virginia, University 
of Washington, and Yale University.  \\

This work was funded with grants from Consejo Nacional de Investigaciones 
Cient\'{\i}ficas y T\'ecnicas and Universidad Nacional de La Plata
(Argentina).

\end{acknowledgements}

%
%
\bibliographystyle{aa}
\bibliography{dSph_HCG44_rev3_arxiv}

\end{document}